\documentstyle[twocolumn,aps]{revtex}

\begin{document}

\title{Diffusion in disordered media as a process with memory}
\author{Michele Vendruscolo$^1$ and Matteo Marsili$^2$}
\address{$^1$ Istituto Nazionale per la Fisica della Materia (I.N.F.M.)\\ and
International School for Advanced Studies (S.I.S.S.A.),
Via Beirut 2-4, 34014 Trieste, Italy}
\address{$^2$Institut de Physique T\'eorique, Universit\'e de Fribourg, CH-1700}

\address{\em (\today)}
\address{~}


\address{
\centering{
\medskip\em
\begin{minipage}{14cm}
{}~~~The problem of a random walk in a disordered media is mapped into
a model of a random walk with memory. The latter model, as opposed
to the former one, does not make reference to a particular realization
of the disorder. The equivalence of the two
models implies that the new model retrieves dynamically a realization
of disorder; the only one which is consistent with its dynamics.
In this new approach to the dynamics in disordered media, effects
of memory, aging and the peculiar localization properties of the
random walker, appear quite natural. 
{}~\\
\medskip
{}~
\end{minipage}
}}

\maketitle


\narrowtext

The dynamics of disordered systems is a very active subject
of research of statistical physics. In non equilibrium
systems, such as driven interface growth \cite{interf} and
charge density waves\cite{cdw}, disorder leads to very
interesting effects as depinning transitions, creep
phenomena and self organization. In out
of equilibrium systems, like spin glasses, aging effects
arise which, at least at a mean field level, has been
related to the lack of time translational invariance and the
failure of fluctuation dissipation relations \cite{kurch}.
The main complication brought by the presence of disorder
is that, in order to compute a physical quantity, apart from the
``dynamic'' average over different stochastic time
evolutions, quenched dynamics requires a second
average over the realizations of disorder.
This, operationally, implies that one has to evolve the
system in several disorder configurations and at the end
average the result over the realizations of disorder.
On one hand, the dynamics explicitely depends on the
particular
realization of the disorder (typically through
transition rates).
On the other, in most systems, one expects the
physical quantities to be self averaging and
therefore to depend weakly on the disorder
configuration.
This situation is rather unsatisfactory, in our
opinion, because only after this second average over
disorder it is possible to appreciate the general
features of the dynamics.
It has recently been pointed out \cite{RTS} that this problem
can be overcome in non equilibrium models based on extreme
dynamics, by appealing to an annealed dynamics (we shall
use this term as opposed to quenched dynamics) which
does not make reference to a particular realization of
disorder. The advantage of this point of view is that
only the average over different stochastic time
evolutions need to be taken: the effective dynamics is
indeed such that the averages over disorder
are taken ``run time'', i.e. at each time step, by the
process itself.
Moreover this approach provides also the
statistical weight of the history of the process, which is
hardly available in dynamics with disorder. The key point,
in the derivation of such annealed dynamics, is that the
future evolution has to be statistically consistent with the
past history. The mathematical translation of this principle
relies on the concept of conditional probability. The
process thus acquire time dependences which naturally
explain the emergence of memory effects in quenched
dynamics. It has also been shown that, from this point of
view, the relation between extremal dynamics and self
organization are a simple consequence of a more general
relation between dynamical processes with memory and self
organization \cite{memo}.

The purpose of this paper is to apply the same
considerations to an equilibrium system. We shall deal
with the simplest such system, i.e. a one
dimensional random walk in random environment. For this
we will derive the exact corresponding annealed dynamics. 
This dynamics, by definition, does not depend on any
particular realization of the disorder. However, as we
shall see, the process has the same statistical
properties. Asymptotically, for large times, the process
singles out a particular realization of the disorder, which
is the only one which is consistent with the past history of
the process. 
A simple generalization of the dynamics with memory
we find, shows that, interestingly enough, 
the disordered dynamics lies on the border line between
random dynamics and
deterministic dynamics. The random walker, in the latter
case will sooner or later localize on some site. Finally 
we shall generalize our arguments to the problem of a
random walk with traps and draw some conclusions.

The random random walk (RRW) on a line is defined by
assigning at each site $i=0,\pm 1,\pm 2,\ldots$ a random
variable $p_i\in [0,1]$ drawn from a distribution 
$P\{p\le p_i<p+dp\}=\phi(p)dp$. The evolution of the 
position $x_t$ of the RRW is
defined by $x_{t+1}= x_t+1$ with probability $p_{x_t}$ and
$x_{t+1}= x_t-1$ otherwise. In spite of its simplicity 
this model has been studied by
many authors as a toy model for localization
\cite{tosatti}, depinning transitions \cite{RRW} and aging
effects \cite{mp}. The most striking
feature is that the diffusion is extremely slow: The
typical size visited by the walker after a time $t$ is
$\delta x\sim (\ln t)^2$. Comparing this result,
originally derived rigorously by Sinai \cite{sinai}, with
the diffusion of a random walk without disorder, $\delta
x\sim  \sqrt{t}$, suggests that disorder has really
dramatic effects on the dynamics. 

In order to introduce our model, let us consider the case
of a uniform distribution $\phi(p)$=1. Imagine to observe
the walker in its motion, without knowing the realization
$\{p_i\}$ of the disorder. The only information available
is what one sees, namely the number $n_{i,t}$ of times that
the random walker has visited site $i$ and the number
$k_{i,t}$ of times in which it has moved from
site $i$ to site $i+1$. 
As we shall now show, it is possible, using this information, 
to describe a RRW even if the values of $p_i$ are not known. 
This is accomplished by observing that the
probability that the number of right jumps $i\to i+1$ is
$k$, given that site $i$ has been visited $n$ times and the
transition probability is $p_i=p$, is simply given
by the binomial distribution
\begin{equation}
P(k|n,p)={n\choose k}p^k(1-p)^{n-k},
\label{1}
\end{equation}
where the notation $P(A|B)$ stands for the probability of
the event $A$, conditional to the occurrence of $B$. Regarding
$k$ as the ``effect'' of the ``cause'' $p$, we can invert
this statistical relation to obtain the probability
$dP(p|n,k)$ that $p\le p_i<p+dp$ given $k$ and $n$. Using
Bayes rule of causes (see \cite{feller} p. 124), it is easy 
to find that $dP(p|n,k)=(n+1)P(k|n,p)dp$. From this we can 
obtain an ``effective'' transition probability 
\begin{equation}
p^{a}_{n,k}=\int dP(p|n,k) p =\frac{k+1}{n+2}
\label{2}
\end{equation}
where the last equality holds for $\phi(p)=1$ (see later).
The content of eq.~(\ref{2}) is that, among all the processes
and all the realizations of the disorder, the probability
that the random walker will jump from site $i$ to site
$i+1$, given that it has made the same jump $k$ times after
the $n$ previous visits, is $p^{a}_{n,k}$.
This is the transition probability which is consistent, in a
conditional way, to the past history of the process. 
The history of the process is in general encoded in the
effective distribution of the variable $p_i$ at time $t$,
which was named run time statistics in \cite{RTS}. In our
case the distribution of $p_i$ is parametrized by only two
numbers $n_i$ and $k_i$, and therefore a direct expression
of the effective dynamics in terms of $k_i$ and $n_i$ only
is possible. The structure of
the memory can be described by placing a Polya urn on each
site \cite{feller}.

The model defined by eq.~(\ref{2}) will be hereafter called
a random walk with memory (RWM). Its evolution is defined
as follows: define on each site $i$ of the lattice two
integer ``dynamical'' variables $n_{i,t}$ and $k_{i,t}$ which count
the number of visits on site $i$ and the number of jumps
$i\to i+1$. At time $t=0$, $n_{i,0}=k_{i,0}=0$ and the
walker is at site $i=0$.  At time $t$, if the random walker
is at site $i$, then with probability
$p^{a}_{n_{i,t},k_{i,t}}$ it will move to site $i+1$ and
$k_{i,t+1}=k_{i,t}+1$. Otherwise the walker moves to site
$i-1$ and $k_{i,t+1}=k_{i,t}$. In either
case  $n_{i,t+1}=n_{i,t}+1$ increases by one. This process, 
by construction, is
expected to reproduce the same results of the RRW with a
random realization of $\{p_i\}$. 
In the RWM, the transition probabilities depend on the
dynamical variables $\{k_{i,t},n_{i,t}\}$ and therefore
evolve in time.
On the contrary, in the RRW, the transition probabilities 
$p_i$ are fixed before the process starts.
The equivalence of the dynamics of the two walkers results
from the fact that each realization of the RWM 
asymptotically singles out
a realization of the disorder, in the sense that
$p^{a}_{n_{i,t},k_{i,t}}\to p_i$ as $t\to \infty$, where $p_i$
is a uniform random number in $[0,1]$. 
This has been explicitly checked in numerical simulations,
but it can also be argued from the distribution
$dP(p_i|n,k)/dp$ of $p_i$. This is indeed sharply peaked 
around the mean value $p^{a}_{n,k}$, with a width of order
$1/\sqrt{n}$. The statistics of the asymptotic value
of $p^{a}_{n,k}$ as $n\to\infty$ can be explicitly shown
to be that of uniform random variables by
analyzing the moments of the effective transition
probability $p_i(n_i)=p^{a}_{n_i,k_i}$. Dropping the $i$
index for the moment, one observes that at the $n-1^{\rm
st}$ visit $p(n-1)^q$, with probability $p(n-1)$ 
increases to $\left[\frac{(n+1)p(n-1)+1}{n+2}\right]^q$ 
while with probability $1-p(n-1)$ it becomes
$\left[\frac{(n+1) p(n-1)}{n+2}\right]^q$. Taking the
average over realizations, leads to a recursion relation for
the moments of $p(n)$ which, with a little algebra, can be
solved to find \begin{equation}
M_q(n)=\langle{p(n)^q}\rangle=\frac{1}{n+1}\sum_{k=1}^{n+1}
\left(\frac{k}{n+2}\right)^q.
\label{mom}
\end{equation}
Note that $M_1(n)=1/2$ for all $n$. Moreover all 
central moments $\langle{[p(n)-\langle{p(n)}\rangle]^q\rangle}$ with $q$
odd vanish identically. For $n\gg 1$, one easily finds
$M_q(n)=(1+q)^{-1}+O(n^{-1})$, i.e. the moments of $p(n)$
tend indeed to those of a uniform distribution in $[0,1]$.
Therefore, the distribution of the transition probabilities,
for a RWM in a box of size $L$ with periodic
boundary conditions, will asymptotically tend to a delta
function around a random value $p_i$ whose
statistics is uniform in $[0,1]$. However, strictly
speaking, even with periodic boundary conditions, the random
walk will never reach a stationary state. This is
reminiscent of systems out of equilibrium. 

Another interesting observation is that one can easily
calculate the probability of a realization of the process,
i.e. of a given history $\{x(\tau):\tau=1,t\}$. This is
indeed given simply by $P\{n_{i,t}\}=\prod_i[n_{i,t}+
1]^{-1}$\cite{note}. Note that to obtain such a quantity in
the RRW, one needs to evaluate it for a given realization
of the disorder and then average over all realizations.

The diffusion law $\delta x\sim (\ln t)^2$ 
can be understood, in the context of the
RWM, with the following argument. First we note that the values
of $k_i$ and $n_i$ on different sites are not independent. 
For example it is easy to check that $t=\sum_i n_i$ and
$x_t=\sum_i (2k_i-n_i)$. In general $n_i=k_{i-1}+n_{i+1}-k_{i+1}$.
In this relation the $k$'s are distributed uniformly between $0$
and the $n$'s. Then, approximately, this relation has the form 
$n_{i+1}\simeq C_i n_i$ with $C_i$ a random variable.
In other words the variable $\ln n_i$ will have the shape of a random
walk over $i$, which means that typically the maximum value of
$n_i$ for $i\in[0,L(t)]$ will be $n_{\max}\sim\exp\sqrt{L(t)}$. 
Since this value will also dominate the sum $\sum_i n_i=t$, we 
can conclude that $L(t)\sim (\ln t)^2$.

One striking feature of the RRW is the lack of time 
translational invariance. It was pointed out \cite{mp}
that two times correlation functions are not functions 
of the difference of the times, as is normally the case, 
but also depend on the ``waiting'' time (i.e. the smallest 
time). This was related in ref. \cite{mp} to the
aging phenomena observed in spin glasses and glasses.
The calculation of $\langle{A_t A_{t+\tau}}\rangle$, where $A_t$ is
any observable, depends only on processes between times $t$ 
and $t+\tau$. If the transition probabilities involved in these
process are constant in time, time translation invariance
follows naturally. The lack of time translational invariance
is no surprise in the RWM, because the transition probabilities 
explicitly depend on the ``waiting'' time $t$. 
This point can be hardly appreciated in the framework of
the RRW, where the transition probabilities are fixed from
the beginning. 
The absence of quenched disorder in the RWM evidences the
fact that aging effects result from local memory effects.
These effects, as shown by the equivalence of the RRW and
RWM, are also present in disordered dynamical systems.

One might wonder what happens if instead of a uniform
distribution one considers a general distribution
$\phi(p)$. It is not difficult to show that all the above
considerations hold the same, apart from the specific form
of the moments and of the distribution of $p_i(n)$. 
Indeed eq.~(\ref{1}) still holds. However when one inverts it to find
the distribution $dP(p|n,k)$ one has to account for the
fact that the probability that $p\le p_i<p+dp$ is
$\phi(p)dp$ with $\phi(p)\ne 1$ in general. 
In practice eq.~(\ref{2}) is slightly modified, but only up
to factors of order $n^{-1}$. For example, if
$\phi(p)=\Gamma(\alpha+\beta) x^{\alpha-1}(1-x)^{\beta-1}/[\Gamma(\alpha)
\Gamma(\beta)]$, one finds  $p^a_{n,k} = (k+\beta)/(n+\alpha+\beta)$. 
Our numerical check of 
the diffusion as a function of $\alpha$ for $\beta=1$ confirm the 
depinning transition for $\alpha>2$ found by Derrida~\cite{d83}.

To address the problem of localization 
we note that on each site the RWM can create a barrier. If the walker
has failed to pass a site after $n$ visits, its probability
to overcome it at the next visit is $p^a_{n,0}=1/(n+2)$. Even though
this probability decreases, it decreases so slowly that
any barrier will sooner or later be overcame. This results
from a straightforward application of the Borel Cantelli
lemma\cite{feller}. It is worth to observe that this behavior 
is the probabilistic counterpart of the ``marginal'' localization 
properties of the RRW\cite{tosatti}.
Indeed it is easy to show, by the same argument, that 
if $np^a_{n,0}\to 0$, as $n\to\infty$ the RWM would surely localize, 
sooner or later on some site. 
This marginality seems to be even stronger as suggested by the following
argument. For any regular distribution $\phi(p)$, we found 
$np^a_{n,0}\to 1$ as $n\to\infty$. Let us therefore generalize
our model by taking 
\begin{equation}
p^a_{n,k}=\frac{k+1}{n+2}+a\sin \left(2\pi\frac{k+1}{n+2} \right).
\label{fx}
\end{equation} 
This describes a generalized symmetric ($p^a_{n,k}+p^a_{n,n-k}=1$) 
random walk with memory. Note that $np^a_{n,0}\to 1+2\pi a$.
We expect that, for $a<0$ the walker localizes, whereas
for $a>0$, for large times, the dynamics becomes that of a
random walker without disorder (i.e. $p_i=1/2$).
This expectation is based on the fact that the 
function $f(x)\equiv p^a_{n,xn}$ seen as a map [i.e.
$x_{n+1}=f(x_n)$] has two stable fixed points (0 and 1) and one
unstable fixed point (in $x=1/2$) in the first case ($a<0$)
while in the second case the stability is reversed (0,1 are
unstable and 1/2 is stable). Our problem is not a map, but
it is similar (it has also randomness). However, numerical 
investigation shows that our expectation is correct.
For $a<0$ the walker localizes, whereas for $a>0$ all
the transition probabilities $p_i\to 1/2$ as $t\to\infty$.
In other words, as shown in fig.~\ref{fig_phi},
the dynamics recovers different
distributions of the disorder in the three cases:
\begin{equation}
\begin{array}{lr}
\phi(p)=\frac{1}{2}\delta(p)+\frac{1}{2}\delta(p-1) &\hbox{for $a<0$}\\
\phi(p)= 1 &\hbox{for $a=0$}\\
\phi(p)=\delta\left(p-\frac{1}{2}\right) &\hbox{for $a>0$}
\end{array}
\end{equation}
From this point of view the case $a=0$ is very peculiar.
It is the only case for which the distribution
which is recovered by the dynamics is continuous.
The case $a<0$ bears some resemblance with
systems, such as the Hopfield model\cite{hop} or folding 
proteins\cite{prot}, where the phase space has a peculiar 
organization and the dynamics ``localizes'' on a particular
low energy state. 

The above model can be generalized straightforwardly to higher
dimensions $d$. This only requires the introduction of $d$
dynamical variables $k_i^{(j)}$, $j=1,\ldots,d$, one for each direction
on each site. An even simpler generalization is the case of a 
$d$ dimensional random walker with random traps: Assign a 
uniform variable $p_i\in [0,1]$ to each site of the lattice.
If the walker is on site $i$ at time $t$, with probability $p_i$
it remains on the same site at $t+1$, and with probability $1-p_i$
it diffuses to one of the neighbor sites. Still we can use 
$p^a_{n_i,k_i}$ for the probability of jumping out of site $i$,
conditional to $n_i$ visits and $k_i$ previous jumps out of
the trap. It is easy to see how
the diffusion law is modified in this case. Indeed, apart from the
fact that the walker can spend a time $n_i>1$ over a given site
before jumping to the next one, the diffusion is the same. This
means that $\delta x^2\sim N$ where $N$ is the number of sites
visited (i.e. the number of jumps). 
This is related to the time $t$ by summing all the times
spent on different sites: $t=\sum_{i=1}^N n_i$. This sum is dominated 
by the large $n_i$ values. 
The probability that the walker has been
trapped for $n_i$ steps on site $i$ is $(n_i+1)^{-1}$. 
The probability that it will jump out of the trap is 
$p^a_{n_i,0}=1/(n_i+2)$. Therefore the distribution of 
$n_i$ is $D(n)=[(n+1)(n+2)]^{-1}$. This means that, for $N\gg 1$,
$t=\sum_{i=1}^N n_i\sim N\ln N$, which yields the diffusion law
$t\sim \delta x^2 \ln \delta x^2$. We checked the logarithmic 
corrections to the diffusion numerically. 
In this case, using the generalized model of eq. (\ref{fx}),
it is easy to find that $D(n)\sim n^{-2-2\pi a}$. Therefore 
for $a>0$, the above argument yields the standard 
diffusion $\delta x^2\sim t$,
whereas for $a<0$ one finds anomalous diffusion $\delta x^2\sim 
t^{1+2\pi a}$. Also in this case, therefore, disorder dynamics
appears to be a borderline case.

In conclusion we have derived and discussed some simple models
of random walks which reproduce the behavior of diffusion
in disordered media {\em without} specifying the disorder.
We have seen that the dynamics itself retrieves a realization
of the disorder with the proper statistical properties. 
Our results may well be used to generate dynamically a random
realization of the disorder in any model with quenched variables.
It is tempting to conjecture that such an algorithm could
provide an alternative to the simulated annealing \cite{simulanneal}
procedure used to find optimal configurations in disordered systems.
The annealing procedure has indeed the drawback that, once the
disorder realization is fixed, the starting configuration of
the dynamical variables may be ``far'' from a reasonably good
optimal state.
Using the above results would instead produce dynamically a realization
of the disorder which is ``consistent'' with the configuration 
of the dynamical variables.


\begin{figure}
\caption{Probability density $\phi (p)$. The solid curve centered in $p=0.5$
is obtained for $a=0.1$ and is reminiscent of a random walk.
The dotted lines are previous stages of simulation. 
The solid curve with two peaks in $p=0$ and $p=1$ refers
to the $a=-0.1$ case where localization takes place.}
\label{fig_phi}
\end{figure}



\begin{thebibliography}{99}

\bibitem{interf} L.-H. Tang and H. Leshhorn, Phys.
                Rev. A {\bf 45}, R8309 (1992); S. V. Buldyrev, A.-L.
                Barab\'asi, F. Caserta, S. Havlin, H. E. Stanley and
                T.Vicsek, Phys. Rev. A { \bf 45}, R8313 (1992).
\bibitem{cdw}   G. Gruner, Rev. Mod. Phys. {\bf 60}, 1129 (1988).

\bibitem{kurch} L. Cugliandolo and J. Kurchan, 
               Phys. Rev. Lett. {\bf 71}, 1 (1993)
\bibitem{RTS}  L. Pietronero and W. R. Schneider, Physica A {\bf 119},
               249-267 (1989); M. Marsili, J. Stat. Phys. {\bf 77}, 733 (1994).
\bibitem{memo} M. Marsili, G. Caldarelli and M. Vendruscolo,
               Phys. Rev. E {\bf 53}, R1 (1996)
\bibitem{tosatti} E. Tosatti, M. Zannetti, and L. Pietronero,
                  Z. Phys. B {\bf 73}, 161 (1988)
\bibitem{RRW} J. P. Bouchaud, A. Comtet, A. Georges and
                P. Le Doussal, Ann. Phys. {\bf 201}, 285 (1990).
\bibitem{mp} E. Marinari and G. Parisi, J. Phys. A: Math. Gen.,
                {\bf 26}, L1149 (1993).
\bibitem{sinai} Ya. G. Sinai, {\em Theory of Prob. and its Appl.}
                {\bf 27}, 256 (1982).
\bibitem{feller} W. Feller: {\em An Introduction to Probability 
                Theory and its Applications}, Ed. J. Wiley \& Sons (1968).
\bibitem{note} It can be shown that $k$ is an integer with an uniform 
                distribution in $0,1,\ldots,n$.
\bibitem{d83}  B. Derrida, J. Stat. Phys. {\bf 31}, 433 (1983).

\bibitem{hop} J. J. Hopfield, Proc. Natl. Acad. Sci. U.S.A., 
              {\bf 79}, 2554 (1982)
\bibitem{prot} P. G. Wolynes, J. N. Onuchic and D. Thirumalai,
               Science {\bf 267}, 1619 (1995)
\bibitem{simulanneal} S. Kirkpatrick, C. D. Gelatt Jr and M. P. Vecchi,
                Science {\bf 220}, 671 (1983).

\end{thebibliography}
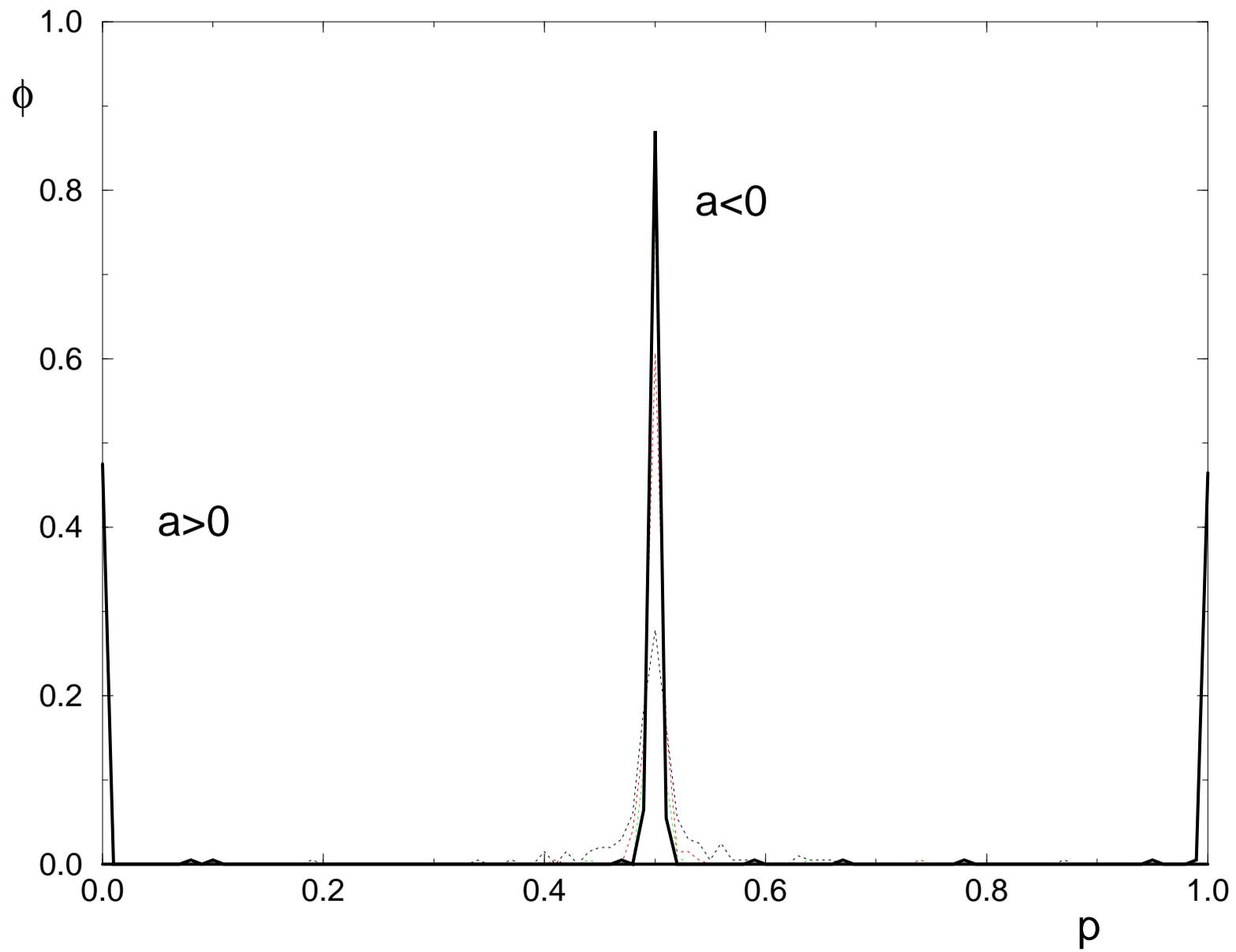
\end{document}